\documentstyle[11pt,osid]{article}

\title{Simulating Ramsey-Type Fringes in a Pulsed Microwave-Driven
Classical Josephson Junction}
\author{Jeffrey E.~Marchese\thanks{$\;$Supported by the UC~Davis Center
for Digital Security, AFOSR grant FA9550-04-1-0171} \\
{\footnotesize\it Department of
Applied Science, University of California, Davis, California 95616, USA}\\[2ex]
Matteo Cirillo\thanks{$\;$Supported by MIUR (Italy) COFIN04} \\
{\footnotesize\it Dipartimento di Fisica,
Universit\`{a} di Roma "Tor Vergata", I-00173 Roma, Italy} \\[2ex]
Niels Gr{\o}nbech-Jensen\thanks{$\;$Supported by the UC~Davis Center
for Digital Security, AFOSR grant FA9550-04-1-0171} \\
{\footnotesize\it Department of
Applied Science, University of California, Davis, California 95616, USA} }

\begin{document}

\maketitle
\begin{abstract}
We present evidence for a close analogy between the nonlinear behavior
of a pulsed microwave-driven Josephson junction at low temperature and the
experimentally observed behavior of Josephson systems operated below
the quantum transition temperature under similar conditions. We specifically
address observations of Ramsey-type fringe oscillations, which can
be understood in classical nonlinear dynamics as results of slow transient
oscillations in a pulsed microwave environment. Simulations are
conducted to mimic experimental measurements by recording the statistics
of microwave-induced escape events from the anharmonic potential well of
a zero-voltage state. Observations consistent with experimentally found
Ramsey-type oscillations are found in the classical model.
\end{abstract}

\section{Introduction}
Macroscopic quantum behavior of Josephson systems has been the focus
of intense research over the past years due to the potential applications
of Josephson technology in quantum information processing
\cite{QCQBMS}. Many experimental
observations have been conducted in order to characterize and understand
the quantum properties of various Josephson device and networks \cite{
Vion_02,
Martinis_02,
Vion_03,
Martinis_03,
Chiorescu,
Claudon_04,
Kutsuzawa,
Wallraff,
Plourde,
Koch,
Simmonds}.
Due to the nature of the Josephson effect it is not easy to make direct
observations of the expected quantum states in the anharmonic potential
of a zero-voltage state Josephson system, so most measurements are
conducted statistically by recording the distributions of transitions
from one system state to another. These transitions are commonly induced
through the controlled application of microwaves and microwave pulses
of frequencies that are commensurate with the expected energy separations
between possible quantum levels of the system.

Some of the key microwave induced features that have been used to illustrate
the quantum properties of Josephson systems are multi-peaked switching
distributions as a function of bias-point in continuously applied microwaves
\cite{Wallraff_03,Jensen_04_1},
observations of Rabi-oscillations \cite{Rabi} in the switching statistics
of systems perturbed by pulsed microwaves
\cite{Martinis_02,Claudon_04,Simmonds}, and the associated
Ramsey-fringes\cite{Ramsey} when several coordinated and parameterized
microwave pulsed are applied sequentially \cite{Plourde,Koch}.
These observations, along with many others, have produced
a large body of insight to the dynamical response of Josephson systems
to microwave perturbations applied at very low temperature, and the acquired
data has been used to interpret the quantum nature of the superconducting
networks under
investigation. However, many of the observations may also be attributed to
classical nonlinear dynamics and statistics. For example, it has been
demonstrated that the multi-peak switching distributions as a function
of varying system bias for a constantly applied microwave field can be
understood as classical high-Q resonances in the nonlinear system
\cite{Jensen_04_1, Jensen_04_2}. Also Rabi-type oscillations in the
switching statistics of Josephson systems perturbed by microwave
pulses have been shown to be consistent with theoretical and
numerical expectations based on a classical model of a Josephson junction,
which exhibit the transients in response to the pulsed perturbations
\cite{Jensen_05, Marchese_05}. It is the purpose of this paper to
demonstrate that Ramsey-type fringes can also be observed in the classical
model when the recipe of the experimental measurements is followed.
We here, for simplicity, illustrate the results with the simplest
possible system, namely a single classical Josephson junction, perturbed
by a bias-current, microwave fields, probe fields, and thermal noise.

\section{The Model}
\begin{figure}
   \setlength{\unitlength}{0.45 cm}
   \begin{picture}(12,12)(0,0)
 \includegraphics{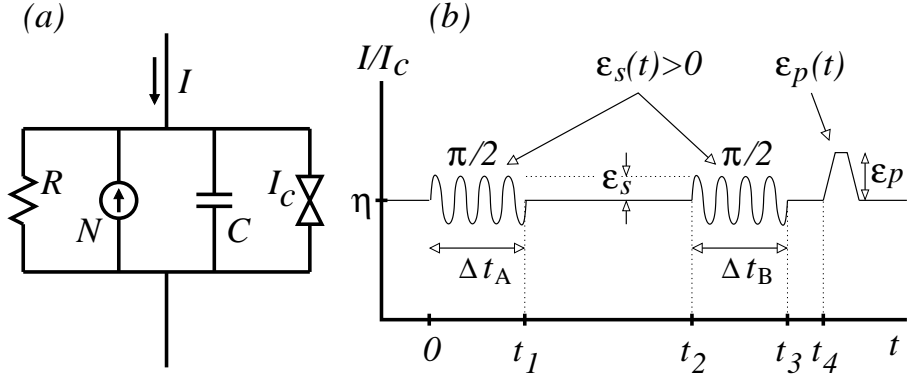}
   \end{picture}
   \vspace{1.0 cm}
   \caption{(a) Sketch of the RCSJ model where
$I/I_c=\eta+\varepsilon_s(t)\sin(\omega_st+\theta)+\varepsilon_p(t)$
and $N=n(t)I_c$. (b) Sketch of the current supply to the junction.}
   \label{fig:figure1}
\end{figure}

The normalized classical equation for a perturbed Josephson junction within
the RCSJ model (see Figure 1a) is \cite{Barone_82},
\begin{eqnarray}
\ddot{\varphi}+\alpha\dot{\varphi}+\sin\varphi & = &
\eta+\varepsilon_s(t)\sin(\omega_st+\theta_s)+\varepsilon_p(t)+n(t) \; ,
\end{eqnarray}
where $\varphi$ is the difference between the phases of the
quantum mechanical wave functions defining the junction, $\eta$
represents the dc bias current, and $\epsilon_s(t)$, $\omega_s$,
and $\theta_s$ represent microwave amplitude, frequency, and
phase, respectively (see Figure 1b). All currents are normalized to the
critical current $I_c$, and time is measured in units of the inverse plasma
frequency, ${\omega_0}^{-1}$, where ${\omega_0}^2=2eI_c/\hbar
C=2\pi I_c/\Phi_0 C$, $C$ being the capacitance of the junction,
and $\Phi_0=h/2e$ is the flux quantum. Tunneling of quasiparticles
is represented by the dissipative term, where
$\alpha=\hbar\omega_0/2eRI_c$ is given by the shunt resistance R,
and the accompanying thermal fluctuations are defined by the
dissipation-fluctuation relationship \cite{Parisi_88}
\begin{eqnarray}
\left\langle n(t) \right\rangle&=&0 \\
\left\langle n(t)n(t') \right\rangle&=&
2\alpha {k_BT \over H_J}\delta(t-t')=2\alpha \Theta\delta(t-t'),
\end{eqnarray}
$T$ being the thermodynamic temperature, $H_J$ is the characteristic
Josephson energy $H_J=I_c\hbar /2e$, and $\Theta$ is thereby defined
as the normalized temperature. A current pulse for probing the state
of the system is represented by $\epsilon_p(t)$.

\section{Simulation Details}

Following the procedure of reported experiments, we record the switching
from the zero-voltage state ($\langle\dot\varphi\rangle=0$) as a
result of applying the external current sketched in Figure 1b. We first
equilibrate the system at a chosen temperature for a given value of $\eta$.
For a randomly chosen, but temporally constant, phase $\theta_s$, a
microwave pulse with frequency $\omega_s$, which is associated with
the natural resonance of the junction at bias point $\eta$, is applied
for a duration of $\Delta t_A$. This duration is chosen to be
$\Delta t_A=\tau_R/4$, where $\tau_R=2\pi/\Omega_R$ is the Rabi-type
oscillation period
\cite{Jensen_05,Marchese_05} for the microwave amplitude $\varepsilon_s$.
This pulse is denoted a $\pi/2$-pulse due to its parameterization based
on Rabi-type oscillations.
Another $\pi/2$-pulse, in phase $\theta_s$ with the first pulse, is applied
at a later time with identical amplitude, frequency, and duration
($\Delta t_B=\Delta t_A$), whereafter a short pulse $\varepsilon_p(t)$
is applied to probe whether the system is in a high or low energy state
after the sequential microwave pulse application. The probe pulse is
parameterized such that a state of relatively high system energy at $t_3$
will result in a likely escape event when $e_p(t)$ is activated, while
a state of relatively low system energy will result in a low probability
of escape when $\varepsilon_p(t)$ is activated. This procedure is repeated
many times, each time with a new realization of $\theta_s$, in order to generate
information about the probability of switching for a given set of parameters,
such as the time delay between the two $\pi/2$-pulses.
\begin{figure}
   \setlength{\unitlength}{0.45 cm}
   \begin{picture}(12,17.0)(0,0)
     \put(-0, -1){\includegraphics{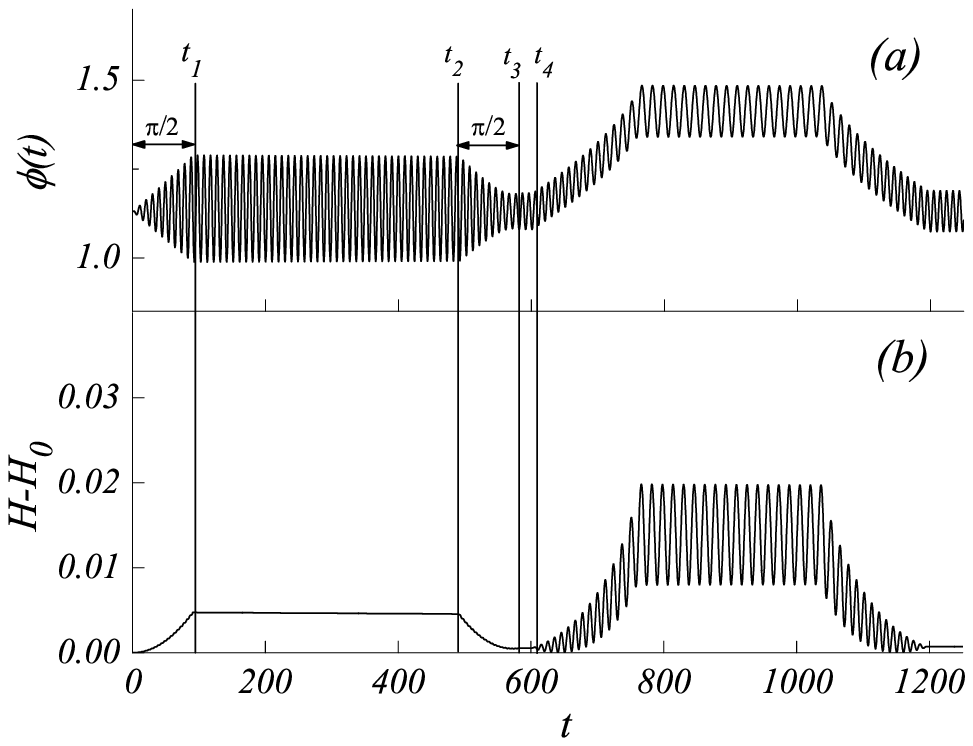}}
   \end{picture}
   \begin{picture}(12,18.0)(0,0)
     \put(-0, -1){\includegraphics{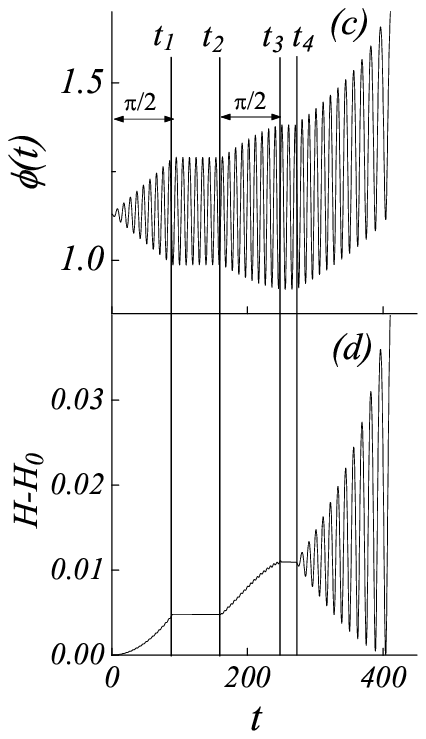}}
   \end{picture}
   \vspace{0.0 cm}
   \caption{Josephson response to the sequential application of two
$\pi/2$ microwave pulses followed by a probe field. (a,c) Josephson phase.
(b,d) System energy as defined by Equations (4) and (5). Parameters are
$\alpha=10^{-4}$, $\eta=0.904706$, $\varepsilon_s=2.17\cdot10^{-3}$,
$\varepsilon_p=8.2\cdot10^{-2}$, and $\Theta=0$. (a,b) Non-switching for
$\Delta t_d=t_2-t_1=400$.  (c,d) Switching for $\Delta t_d=t_2-t_1=70$.}
   \label{fig:figure2}
\end{figure}
Figure 2 illustrates typical examples of the events described above for
$\Theta=0$. Specific
parameter values are: $\alpha=10^{-4}$, $\varepsilon_s=2.17\cdot10^{-3}$,
$\varepsilon_p=8.2\cdot10^{-2}$, $\omega_s=\omega_l$,
and $\eta=0.904706$, which results
in a linear resonance frequency of $\omega_l=\sqrt[4]{1-\eta^2}=0.652714$.
$\pi/2$-pulses are applied in the
intervals $t\in\,[0;t_1=\Delta t_A]$ and $t\in\,[t_2;t_3=t_2+\Delta t_B]$ with
$\Delta t_A=\Delta t_B=90.75$ time units. This value is obtained from
the corresponding Rabi-type oscillations reported in \cite{Marchese_05}.
After the second $\pi/2$-pulse, a short delay of $\Delta t=20$ is
allowed before the probe pulse $\varepsilon_p(t)$ is initiated at $t_4$.
The probe has linear rise and fall times of
160 normalized time units with a constant value interval of 275 time units.
We show the system response to the microwave perturbations in terms of
both phase $\varphi$ and the normalized energy $H-H_0$, which is defined by
\begin{eqnarray}
H & = & \frac{1}{2}\dot\varphi^2+1-\cos\varphi-\eta\varphi \\
H_0 & = & 1-\sqrt{1-\eta^2}-\eta\sin^{-1}\eta \; .
\end{eqnarray}
Figures 2a and 2b show the response for $\Delta t_d=400$. It is
evident that the first $\pi/2$-pulse elevates the system energy by
leaving the system in a phase-locked state at $t=t_2$ with energy
$E(t_1)$. At the onset of the second $\pi/2$-pulse, a significant phase-slip
between the oscillation of $\varphi$ and the microwave has developed
during the interval $[t_1;t_2]$, and the second microwave pulse therefore
decreases the energy of the Josepshon system. The system is left at $t=t_3$
with energy $E(t_3)<E(t_1)$, which, when the probe field is
applied at $t_4$, results in only temporary energy increase while the
probe is applied; this is not enough to make the system switch into a
non-zero voltage state, which can be seen from the fact that $E(t)\rightarrow0$
for large $t$).
For comparison we show Figures 2c and d for $\Delta t_d=70$. We here observe
the same initial behavior, but the application of the second
$\pi/2$-pulse leaves the system in a relatively high energy state
$E(t_3)>E(t_1)$, since the phase-slip between the junction and the
microwave field is minor in this case. Consequently, the junction phase
$\varphi$ switches out of the bound state when the probe pulse is applied
at $t_4$. This is recognized by the diverging energy for $t>t_4$.

\section{Ramsey-Type Fringes}
\begin{figure}
   \setlength{\unitlength}{0.45 cm}
   \begin{picture}(12,15)(0,0)
     \put(-0, -1){\includegraphics{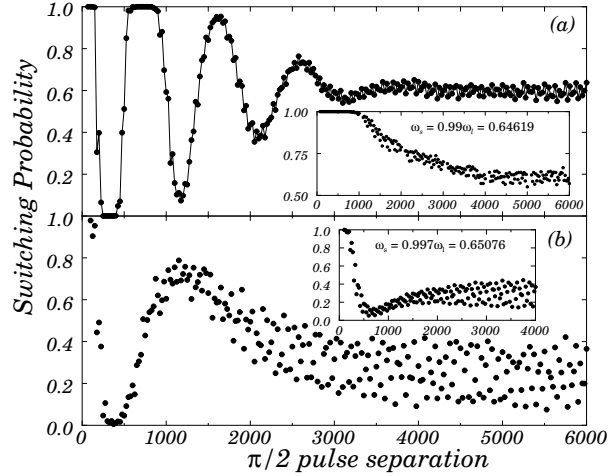}}
   \end{picture}
   \vspace{0.0 cm}
   \caption{Switching probability as a function of $\pi/2$-pulse separation
for $\Theta=2\cdot10^{-4}$. Each point represents
statistics of 2,500 events. Parameters
are as in Figure 2, unless otherwise listed. (a) $\alpha=10^{-4}$. Visible
Ramsey-type fringe oscillation for
$\omega_s=\omega_l=\sqrt[4]{1-\eta^2}=0.652714$. Inset shows vanishing
Ramsey-type fringe frequency near $\omega_s=0.990\omega_l$.
(b) same as (a), except for $\alpha=10^{-3}$.
Inset shows vanishing
Ramsey-type fringe frequency near $\omega_s=0.997\omega_l$.}
   \label{fig:figure3}
\end{figure}

Ramsey-type fringes in the switching probability as a function of
$\pi/2$-pulse separation can be directly observed in Figure 3, where we have
shown the variation of the switching probability for $\Theta=2\cdot10^{-4}$,
calculated as averages of 2,500 events for different values of $\alpha$ and
$\omega_s$. Figure 3a clearly shows a distinct frequency, which we will
name the Ramsey-type fringe frequency $\Omega_F$, for
$\omega_s=\omega_l=\sqrt[4]{1-\eta^2}=0.652714$ and $\alpha=10^{-4}$.
The inset shows that the fringe frequency depends on the applied microwave
frequency such that $\omega_s=0.990\omega_l$ results in $\Omega_F\approx0$.
Similar behavior is observed for larger dissipation parameter $\alpha=10^{-3}$
shown in Figure 3b. Clearly, the fringe-frequency is different for
$\omega_s=\omega_l$, compared to Figure 3a, and in this case
$\omega_s=0.997\omega_l$ results in $\Omega_F\approx0$.

\begin{figure}
   \setlength{\unitlength}{0.45 cm}
   \begin{picture}(12,20.0)(0,0)
   \put(-0, -1){\includegraphics{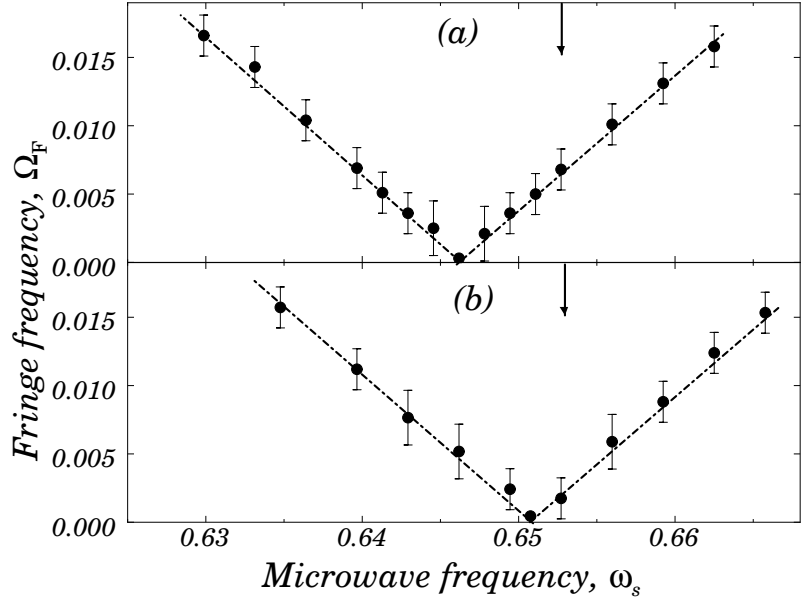}}
   \end{picture}
   \vspace{-1.0 cm}
   \caption{Ramsey-type fringe frequency $\Omega_F$ as a function of
applied microwave frequency for two different dissipation parameters:
(a) $\alpha=10^{-4}$, and (b) $\alpha=10^{-3}$. Other parameters are
as in Figure (3). Arrows indicate the driving frequency for which
   $\omega_s = \omega_l$. Dashed lines indicate a slope of $\pm$1.}
   \label{fig:figure4}
\end{figure}

A number of simulations of the Ramsey-type fringe frequency for
different microwave frequencies have been conducted, and the
results are summarized in Figure 4, where the fringe frequency
$\Omega_F$ is shown as a function of microwave frequency
$\omega_s$ for the two different dissipation parameters mentioned
in Figure 3. The vertical arrows indicate the linear resonance
frequency $\omega_l=\sqrt[4]{1-\eta^2}$. The simulation data are
represented by markers with error bars. The frequencies are read
from figures like Figure 3. For low $\Omega_F$, less than one
wavelength of fringe oscillation may be visible and the
oscillation may be somewhat distorted (see Figure 3b). In such
cases we measured the time from the first trough (peak) to the
next peak (trough) and used the half period to determine the
frequency. In the case of saturation such as Figure 3a, we measured
the period of the first regular waveform. These issues also account for
some of the distortion in Figure 4.

We observe the distinct "V-shape" signature of the Ramsey-type
frequency $\Omega_F(\omega_s)$ \cite{Plourde} with a slope very close to
$\pm$unity (lines with slope $\pm1$ are shown in Figure 4 along
with the simulation data), and $\Omega_F\approx0$ for microwave
frequencies close to $\omega_l$. It is noticeable that the
characteristic microwave frequency for which $\Omega_F=0$
increases with $\alpha$, and we further notice that this
characteristic frequency is smaller than $\omega_l$. While we have
found no experimental data for comparison, future experiments may
want to address this observation.

\section{Conclusions}
Like Rabi-type oscillations and resonant multi-peak switching distributions,
Ramsey-type fringes have been used to identify and analyze expected
macroscopic quantum behavior of Josephson systems. In this paper we have
addressed the comparable classical system through direct simulations
of the well-established nonlinear classical model equation, which is
driven according to the recipe prescribed by the experimental reports
of Ramsey-fringes. Our results show that even the simplest possible
classical Josephson model, the single RCSJ model, clearly exhibits the
features of Ramsey-type fringes when the model is used to simulate
a low temperature, low dissipation system. A detailed theoretical
analysis of classical Ramsey-type fringes \cite{Marchese_06}
can be developed in close
analogy with, and extention of, the analysis of phase-locking transients
that have been found to be responsible for, e.g., Rabi-type oscillations
in pulsed microwave-driven classical Josepshon systems
\cite{Jensen_05,Marchese_05}.

\section{Acknowledgment}
We are grateful to A.~V.~Ustinov for several useful discussions.

\end{document}